\documentclass[twocolumn]{aastex631}

\usepackage{hyperref}
\hypersetup{
    unicode=false,          
    pdftoolbar=true,        
    pdfmenubar=true,        
    pdffitwindow=false,     
    pdftitle={Certificate},    
    pdfauthor={Prof. Dr. Buson Sara},     
    pdfsubject={Origin of astrophysical high-energy Neutrinos},   
    pdfcreator={Prof. Dr. Buson Sara},   
    pdfproducer={Prof. Dr. Buson Sara},  
    pdfkeywords={Blazars} {Neutrinos} {IceCube}, 
    pdfnewwindow=true,      
}

\usepackage{sidecap}
\usepackage{xcolor}
\usepackage{hhline,soul}
\usepackage{booktabs,tabulary}
\usepackage[flushleft]{threeparttable}
\usepackage{wrapfig}

\newcommand{\TXS}{TXS~0506$+$056}

\graphicspath{{./}{./}}
\shorttitle{Beginning a journey across the Universe}
\shortauthors{Buson et al.}

\begin{document}

\title{Beginning a journey across the Universe: the discovery of extragalactic neutrino factories}

\author[0000-0002-3308-324X]{Sara Buson}
\affiliation{Lehrstuhl f\"ur Astronomie, Universit\"at W\"urzburg, Emil-Fischer-St. 31, W\"urzburg, 97074, Germany}
\email{sara.buson@uni-wuerzburg.com, andrea.tramacere@unige.ch}

\author[0000-0002-8186-3793]{Andrea Tramacere}
\affiliation{Department of Astronomy, University of Geneva, Ch. d'\`Ecogia 16, Versoix, 1290, Switzerland}

\author[0000-0003-2497-6836]{Leonard Pfeiffer}
\affiliation{Lehrstuhl f\"ur Astronomie, Universit\"at W\"urzburg, Emil-Fischer-St. 31, W\"urzburg, 97074, Germany}

\author[0000-0003-4519-4796]{Lenz Oswald}
\affiliation{Lehrstuhl f\"ur Astronomie, Universit\"at W\"urzburg, Emil-Fischer-St. 31, W\"urzburg, 97074, Germany}

\author{Raniere de Menezes}
\affiliation{Lehrstuhl f\"ur Astronomie, Universit\"at W\"urzburg, Emil-Fischer-St. 31, W\"urzburg, 97074, Germany}

\author[0000-0002-2515-1353]{Alessandra Azzollini}
\affiliation{Lehrstuhl f\"ur Astronomie, Universit\"at W\"urzburg, Emil-Fischer-St. 31, W\"urzburg, 97074, Germany}

\author[0000-0002-6584-1703]{Marco Ajello}
\affiliation{Department of Physics and Astronomy, Clemson University, Kinard Lab of Physics, Clemson, SC 29634-0978, USA}

\accepted{for publication in ApJL, June 28, 2022}

\begin{abstract}

Neutrinos are the most elusive particles in the Universe, capable of traveling nearly unimpeded across it. 
Despite the vast amount of data collected, a long standing and unsolved issue is still the association of high-energy neutrinos with the astrophysical sources that originate them. 
Amongst the candidate sources of neutrinos there are blazars, a class of extragalactic sources powered by supermassive black holes that feed highly relativistic jets, pointed towards the Earth. Previous studies appear controversial, with several efforts claiming a tentative link between high-energy neutrino events and individual blazars, and others putting into question such relation. 
In this work we show that blazars are unambiguously associated with high-energy astrophysical neutrinos at unprecedented level of confidence, i.e. chance probability of $6 \times 10^{-7}$.
Our statistical analysis provides the observational evidence that blazars are astrophysical neutrino factories and hence, extragalactic cosmic-ray accelerators.

\end{abstract}


\section{Introduction} \label{sec:intro}
Cosmic rays are charged particles of energies up to $10^{20}$eV, far higher than the most powerful human-attained particle accelerator, i.e Large Hadron Collider (LHC). The nature and origin of these particles arriving from deep outer space remain elusive and represent a foremost challenge for the astroparticle and astrophysics fields. Cosmic rays' birthplaces  generate other  particles, neutrinos and $\gamma$ rays among them. Unlike $\gamma$ rays, astrophysical neutrinos are solely created in processes involving cosmic-ray acceleration, making them unique smoking gun signatures of a cosmic-ray source \citep{meszaros2017astrophysical}.
In 2013, the IceCube Collaboration reported the discovery of a diffuse flux of astrophysical neutrinos in the $\gtrsim 100$ TeV to 10 PeV energy range \citep{icecube2013,IC_north_hard_spectrum:2016}. The origin of this diffuse flux is probably extragalactic but has still to be ascertained.

Amongst the candidate sources of high-energy neutrinos there are blazars\footnote{For the definition of blazar adopted here we refer the reader to the 5th Roma-BZCat catalog \citep{massaro20155th}.}, a class of extragalactic sources powered by supermassive black holes harbored at the center of their host galaxies \citep{Hillas1984,Winter:2013,Padovani_simplified:2015,Palladino:2019}. Blazars efficiently convert the gravitational energy of accreting gas into kinetic energy of highly relativistic jets, pointed towards the Earth \citep{Padovani2017}. 
In 2017, the potential association \citep{icecube2018_TXS_flaring,aartsen2018_neutrino_prior_to_TXS} of the $\gamma$-ray bright blazar \TXS\ with putative neutrino emission (chance probability at the $\gtrsim 10^{-4}$ level) has put forward $\gamma$-ray blazars as promising neutrino point-sources, hence cosmic-ray accelerators \citep{Padovani_TXS:2018,Gao:2019,Oikonomou:2021,Keivani:2018,Murase_blazar_flares:2018}.
Further efforts extensively pursued the search for a link between high-energy neutrinos and blazars leading to a large debate, with claimed associations (chance probability $\gtrsim 10^{-3}$) between blazars and high-energy neutrinos \citep{padovani2016extreme,kadler2016coincidence,plavin2021directional,Hovatta:2021,IceCube_AGNcores:2021}, as well as contrasting findings \citep{IceCube2017_2LAC,Yuan:2020}.
Previous studies were hampered by employing a sample of blazars selected according to the objects' electromagnetic properties in a preferential energy band. 
Besides, most searches rely on the assumptions of a correlation between the $\gamma$-ray/neutrino emission \citep{Hooper:2019,giommi2020dissecting,Oikonomou:2019,Garrappa:2019,Franckowiak:2020}, often implying that the majority of the observed $\gamma$ rays originate from the same emission region of neutrinos. 
As shown by several theoretical studies \citep[e.g.,][]{Murase2016_gamma_dark,Reimer2019,IceCube2017_2LAC} and observational constraints \citep{IceCube2017_2LAC,Yuan:2020}, however, a bright GeV $\gamma$-ray emitting blazar can unlikely be at the same time an efficient (co-spatial) producer of high-energy neutrinos.

In this work we overcome the limitations of the previous searches employing the largest  available neutrino dataset optimised for searches of point-like sources and an homogeneous clean sample of the blazar population. The paper is organised as follow: Section \ref{sec:working_hyp} lays out the working hypothesis,  Section \ref{sec:neutrino_data} presents the neutrino data and  Section \ref{sec:bzcat} the blazar sample, Section \ref{sec:cross-corr} describes the statistical analysis and results, Section \ref{sec:pevatrons} and \ref{sec:conclusions} present the discussion and conclusions.

\section{Working Hypothesis}\label{sec:working_hyp}
Blazar theoretical models predict an emerging neutrino spectrum to be hard in the IceCube energy band, with an emission that follows a powerlaw with index $\lesssim-2$ and peaks at $\gtrsim1$~PeV energies \citep[e.g.][]{Mannheim1993,Stecker:2013,Dermer:2014,Murase_external_fields:2014,Petropoulou:2015,Padovani_simplified:2015}. The bulk of the blazars' neutrino emission should reside at energies $\gtrsim1$~PeV. Besides invoking theoretical models in support of this hypothesis, it is demanded by observational constraints such as the IceCube collaboration stacking limit on $\gamma$-ray blazars \citep{IceCube2017_2LAC} that already excludes a substantial ($<27\%$) contribution from this population in the $\sim10$~TeV~/~100~TeV energy range for an emerging neutrino soft spectrum ($\propto E^{-2.5}$). The limit relaxes to 40\% and 80\% when assuming a hard-spectrum, e.g. a powerlaw spectrum $\propto E^{-2.0}$, compatible with the IceCube diffuse flux measured above $\sim 200$ TeV \citep{IC_north_hard_spectrum:2016}. Similar conclusions are drawn by independent, complementary studies \citep[e.g.][]{Yuan:2020}.

Motivated by these primers one may foresee a correlation between blazars and astrophysical neutrinos, especially those of the highest observable energies ($\gtrsim100$~TeV). The IceCube Observatory is sensitive to different astrophysical neutrino energy ranges in the Southern and Northern celestial hemispheres. Given its location at the Geographic South Pole, Earth's opacity hampers the detection of the highest-energy astrophysical neutrinos from the Northern hemisphere; for $\gtrsim100$~TeV neutrino energies the effect starts to be important at $\delta \sim 30^{\circ}$. Therefore, the data collected for the Northern sky are best capable of probing the TeV/sub-PeV range while the Southern data are most sensitive to astrophysical neutrino fluxes in the PeV$-$EeV range \citep[][see also next Section \ref{sec:neutrino_data}]{IceCube_extending50dec:2009}.
Since this work aims testing the hypothesis of blazars as high-energy neutrino emitters, we focus our search on the Southern hemisphere first which provides the most promising discovery ground. A forthcoming publication will address the expansion of this investigation to neutrinos observed at the lower energies ($\lesssim100$~TeV).

\section{The Neutrino Dataset}
\label{sec:neutrino_data} 
The IceCube collaboration has publicly released an all-sky neutrino map encompassing 7 years of observations recorded with the IceCube Neutrino Observatory between 2008-2015 \citep{IceCube7y:2017}. The event sample used to produce this map includes more than 700\,000 individual events and was developed to give optimal performance for the identification of point-sources emitting neutrinos at TeV energies and above. 
To this aim, the individual neutrino events were analyzed using an unbinned likelihood formalism which takes advantage of the good angular reconstruction and large statistics of the dataset \citep{IceCube7y:2017}. The all-sky map produced with such analysis encodes the information about the 7-year time-integrated neutrino emission from steady point-source neutrino emitters.

The IceCube map provides for each direction of the sky at declination $ \mid \delta \mid\; < 85^{\circ}$ a local  probability ($p$-value), in a grid of $0.1^{\circ} \times 0.1^{\circ}$ pixels. The local p-value represents the level of clustering in the neutrino data, i.e. a measure of the significance of neutrino events being uniformly distributed. They are not to be confused with the statistical p-values derived from our correlation study presented in the following. 
In our analysis we adopt the negative logarithm of the provided local p-value, defined as $L = - log(p$-value). The neutrino $L$ values range from 0 to 5.9 and are based on the likelihood of the model assuming an astrophysical source with power-law spectrum in a given direction of the sky. 
Our work treats the local $L$ values as an inference of the direction-dependent neutrino emission, i.e. larger values of $L$ imply a higher probability that a genuine astrophysical signal is responsible for the spatial clustering of neutrino events in such direction of the sky.

Similarly to the strategy used by the IceCube collaboration, we divide the neutrino datasets at the horizon in two regions: the northern sky with declination $85^{\circ} > \delta \geq -5^{\circ}$ and the southern sky with $-85^{\circ} < \delta < -5^{\circ}$. This is motivated by the different energy ranges, background characteristics and analysis techniques applied to the event datasets \citep{IceCube7y:2017}. 

Among differences, in the northern sky data the IceCube collaboration has applied an algorithm to select events which is optimized in the search of point-like sources sensitive to both hard and soft or cutoff signal energy spectra, with power-law spectra $\propto E^{-2}$ and $\propto E^{-2.7}$, respectively, and allow an energy threshold as low as few TeVs. For the southern sky data, the final event rate is optimized to yield the best sensitivity and discovery potential for an $\propto E^{-2}$ spectrum and a lower energy threshold of $\sim 100$~TeV was adopted \citep{IceCube7y:2017}, optimizing the sensitivity and discovery potential for hard-spectrum neutrino signals in the PeV range. The spectral information for the sky map are not available. 
Due to the differences in the background characteristics and analysis procedures between the northern and southern neutrino data, these may be considered different datasets as well as they may be sensitive to multiple populations of astrophysical sources dominating at different energy ranges and/or with different spectra. 
As a matter of fact, the IceCube diffuse neutrino spectrum indicates a spectral break and/or an additional soft-spectrum astrophysical component at lower energies \citep[e.g.][]{IC_north_hard_spectrum:2016}. Contextually, a soft-spectrum (powerlaw index $=3.4$) cluster of neutrinos has been reported consistent with the Seyfert II galaxy NGC~1068 \citep{IceCube10y:2020}.
Given the substantial differences in the  properties and energy ranges of the two datasets, to test the working hypothesis introduced in Section \ref{sec:working_hyp} in the following we focus on the Southern hemisphere dataset.


\section{The Blazar Sample}
\label{sec:bzcat}
To search for astrophysical counterparts to the neutrino spots we use the 5th data release of Roma-BZCat catalog \citep[5BZCat,][]{massaro20155th}. Although it does not contain all existing blazars in the Universe, it is a thorough compilation of 3561 objects of either confirmed or highly likely blazar-nature from the literature where each object has been individually inspected to reduce the possibility of including non blazar sources in the catalog. It has no preferred selection toward a particular wavelength or survey strategy, and offers an homogeneous sample of the blazar population. These are major key differences with respect to previous studies that utilised large samples of blazars selected based on well-defined observational or intrinsic properties \citep{IceCube2017_2LAC,giommi2020dissecting,Resconi2017,plavin2021directional,Hovatta:2021}.

Each 5BZCat object has a detection in the radio band and a complete spectroscopic information available. The only exception are 92 objects that lack an optical spectrum and are thus listed as ``BL Lac candidate". We exclude the latter to maximize the genuineness of our blazar sample.
Besides, due to the deficit of 5BZCat sources near the Galactic plane (given by an observational selection effect where extinction from dust in the Milky Way worsen the sensitivity of observations), we discard from our statistical analysis 5BZCat sources located at Galactic latitudes $ \mid b \mid\; < 10^{\circ} $. We anticipate that a consistent cut is applied in the neutrino sample (see Section \ref{subsec:neutrino_data}). Furthermore, we discard objects at $ \mid \delta \mid\; > 85^{\circ}$, since neutrino data are not available for regions close to the poles. Dividing the blazar sample according to $\delta < -5^{\circ}$, our final southern sky sample hosts 1177 objects.

\section{Statistical analysis procedure}
\label{sec:cross-corr}
No population of high-energy neutrino point-sources has been identified at high confidence at present preventing us from making an useful \emph{a priori} choice for the optimal $L$ parameter that maximizes the astrophysical component in the neutrino data \citep{icecube2013,Icecube:2014,Icecube_north:2015,IC_north_hard_spectrum:2016,IceCube10y:2020,South_IC_Antares_combined:2016}. Similarly, the systematic uncertainties related to the neutrino dataset are not provided.
Given the limited knowledge to guide the selection of the neutrino data, we compute the degree of the blazar/neutrino correlation as a function of two parameters, i.e. the local probability of a neutrino spot to be astrophysical $L_{\rm min}$ and the association radius $r_{\rm assoc}$. These are \emph{a posteriori} cuts. The estimate of the final post-trial statistical significance will incorporate the effect of the scan over the data, as explained in the following sections. 
We note that the blazar catalog incompleteness may have some impact upon the estimate of the strength and optimal set of parameters that drive the blazar/neutrino correlation. In particular, it may weaken the measured strength of a true correlation \citep{AugerAGN:2008}.

\subsection{Neutrino spot sample}
\label{subsec:neutrino_data}
We identify the most-likely location of a putative neutrino source as the pixel-map, i.e. spot, with the highest value of $L$ above a given choice of $L_{\rm min}$. We adopt the Right Ascension ($\alpha_{hs}$) and declination ($\delta_{hs}$) of the pixel-maps as the fiducial sky positions of the spots. We require the separation between neutrino spots to be an angular distance of at least $1^{\circ}$ from each other \citep[similarly to e.g.][]{IceCube7y:2017,IceCube8yNorth:2019}, where $1^{\circ}$ is the upper limit of the systematic error reported by the IceCube collaboration \citep{icecube2013,IceCube_Moon:2014,aartsen2014searches}. 
The choice of $L_{\rm min}$ is a trade-off between the expected astrophysical signal and trials. It is reasonable to assume that for a large majority of the sky locations tested with the likelihood fit, the clustering in the neutrino data is driven by non-astrophysical components due to background fluctuations \citep[][]{IceCube7y:2017}. Therefore, we limit our analysis to the neutrino spots with the highest $L$, i.e. those with pre-trial local significance $> 3.5 \sigma$ Gaussian equivalent. From the southern-sky map we select spot sub-samples defined by a minimum value of $L_{\rm min}$, with $L_{\rm min}$ ranging in [3.5, 4.0, 4.5]. These sub-samples count 55, 21 and 10 spots and provide us with a trade-off between the astrophysical signal and background contamination while at the same time limiting the number of trials. Applying the Galactic plane cut, the final neutrino spot sub-samples $L3.5$, $L4.0$, $L4.5$ contain 44, 19 and 9 spots at $ \mid b \mid\; > 10^{\circ}$, respectively. 

The $L_{\rm min} = 4.5$ threshold coincides with a $2 \sigma$ (pre-trial) tension from background expectation and is close to the largest deviation (above background) of $L_{\rm min} = 4.66$  observed in the all-sky hotspot population analysis carried out by the IceCube\footnote{Note that upon correcting for the large number of trials in the all-sky analysis, the excess is not statistically significant \citep{IceCube7y:2017,Hooper:2019,Smith:2021}.} collaboration \citep[][]{IceCube7y:2017}. The excess in the neutrino clustering data around $L\sim4$ is present also in the one-year dataset and three-year dataset analysis \citep[][]{Hooper:2019,Smith:2021}
and could indicate a near-threshold point-source astrophysical population.
%

\subsection{Positional cross-correlation analysis}
The positional cross-matching strategy employed in the following has been widely used in cross-correlation studies  \citep{Finley:2004,AugerAGN:2008,Resconi2017,plavin2021directional,Hovatta:2021}. Similarly to previous studies \citep{IceCube_Auger_TA:2016,padovani2016extreme, giommi2020dissecting}, we keep fixed the IceCube hotspot positions which are known to be not uniformly distributed. We simulate $10^8$ Monte Carlo (MC) catalogs by randomizing the blazar positions, preserving both the total number and the spatial distribution of the blazars, as confirmed by Kolmogorov-Smirnov tests applied on the spatial distributions. This approach preserves large-scale patterns present in the blazar sample, at the same time allowing us to have a representative set of random cases (Appendix \ref{sec:appendix_statistics} addresses the robustness and impact of the randomization strategy on the results).
To gauge the significance of any excess, the pre-trial p-value is computed by inferring for each angular scale scanned the fraction of simulations having more matches than the real data. We retrieve the distance for which this fraction is minimized, the minimum value of this fraction being the pre-trial p-value. Finally, the post-trial p-value is calculated as the fraction of simulations that, following a similar analysis, would lead to a smaller pre-trial p-value than what observed in the data.

The optimal association radius between the neutrino spots and blazars is driven by the positional uncertainty of the neutrino data. The median angular resolution of the neutrino datasets ranges from $\lesssim 0.4^{\circ}$ to $\lesssim 0.7^{\circ}$ for events with energy-proxy of $\sim100$~TeV and improves with increasing  energies being below $0.6^{\circ}$ at PeV energies \citep{IceCube3y:2013,IceCube7y:2017}. Events with higher energies are expected to be most effective in identifying astrophysical signals with hard spectra in the PeV range. Hence, we test a range of association radii $r_{assoc}$ from $0.4^{\circ}$ to $0.7^{\circ}$.

We perform a grid search approach where we scan the $\{L_{\rm min},r_{\rm assoc}\}$ space, with  $L_{\rm min}$ ranging  in  [$3.5,4.0,4.5$], and $r_{assoc}$ ranging in [$0.4^{\circ}:0.7^{\circ}$] with a step of $0.05^{\circ}$. 
For each set of $(L^{i}_{\rm min},r^{j}_{\rm assoc})$,  we match the positions of the 5BZCat objects with the positions of the neutrino spots. The number of real matches constitutes our test statistics, we denote it as TS$_{\rm astro}^{L_i,r_j}$. Then, we estimate the pre-trial p-value, $p^{L_i,r_j}_{\rm pre}$, for the set of parameters following these steps:
\begin{itemize}
    \item We generate $N_{\rm MC}=10^8$ MC catalogs of blazars randomly shifting the sky position of the original 5BZCat sources by a random value between $0^{\circ}$ and $10^{\circ}$, applying the same cuts as for the real 5BZCat sample, preserving both the total number and the distribution of our blazar sample. 
    In the randomization we ensure that the angular distribution of the catalog is preserved performing Kolmogorov-Smirnov tests on the distributions of the mock Right Ascensions and declinations. The mock-catalog distributions obtained in this manner preserve the statistical properties of the original distribution, e.g. their correlation and non-uniformity across the sky.
    \item We apply the matching procedure between the MC catalogs and neutrino spots. Similarly to what done for the experimental data, for the mock catalogs we allow only one mock-source to be associated with the same neutrino spot, and viceversa. The statistical significance evaluated in this way represents the most conservative approach. However, in reality a neutrino spot could originate from the cumulative neutrino emission of distinct astrophysical sources located at a distance smaller than the instrumental spatial resolution.
    \item The matching procedure yields a distribution of MC test statistics TS$_{k}^{L_i,r_j}$, with $k\in[1,N_{\rm MC}]$.  We derive the chance probability of obtaining a test statistic value equal or higher than the one observed for the real data following \citet{Davison:1997}:
    
    \begin{equation}
        p = \frac{M+1}{N+1}
    \end{equation}
    
    where $M$ is the number of random MC samples with a test statistic equal or larger than in the real data, and $N=N_{\rm MC}$ is the total number of random MC samples generated. This is the pre-trial p-value, $p^{L_i,r_j}_{\rm pre}$, and defines the probability of a chance coincidence following Eq.~1 for the given real set of parameters $(L^{i}_{\rm min},r^{j}_{\rm assoc})$.
\end{itemize}

%
%
\begin{figure}
  \centering
    \includegraphics[scale=0.45]{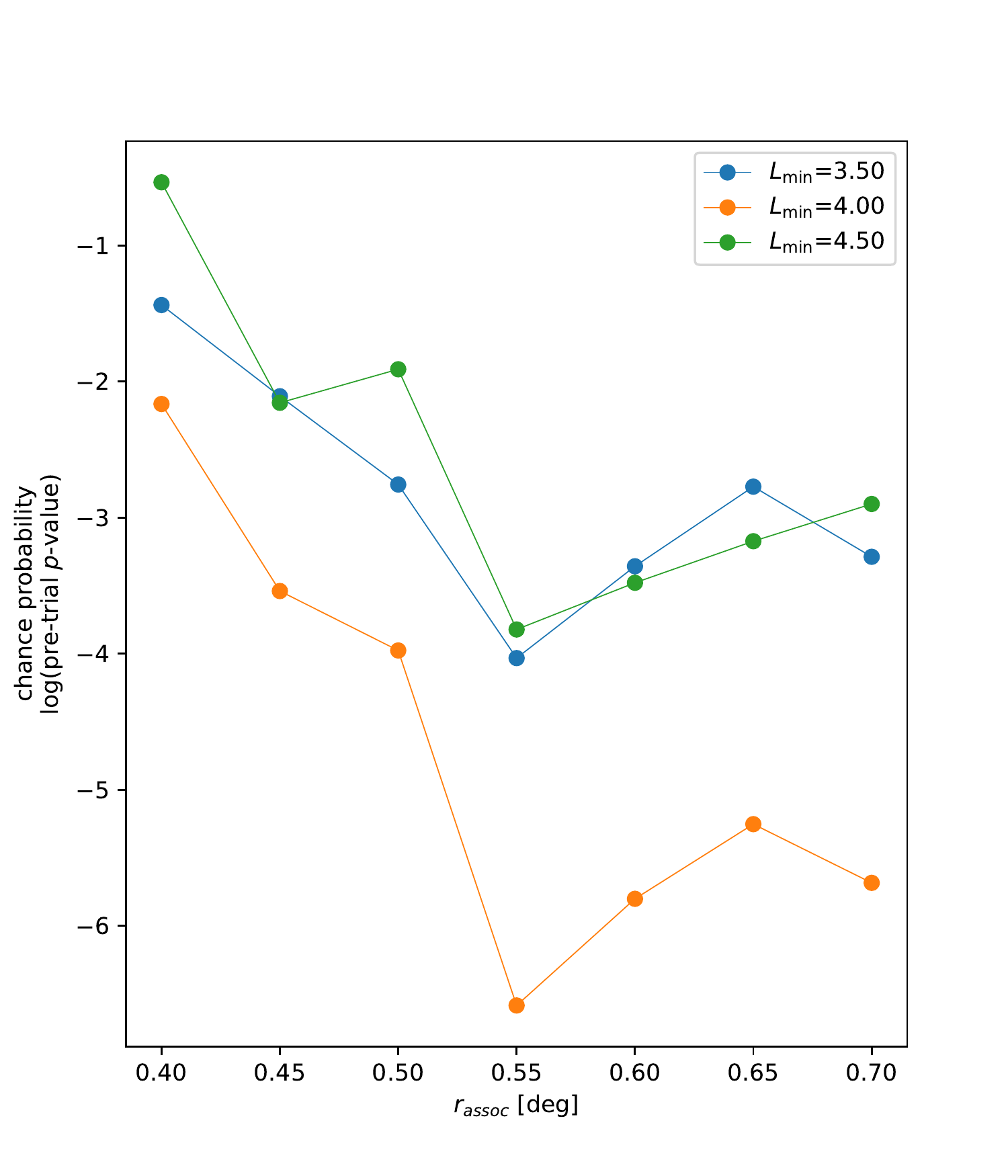} 
        \caption{Pre-trial p-value for the blazar/neutrino correlation as a function of the association radii $r_{\rm assoc}$, for neutrino dataset with $L_{\rm min}=[3.5,4.0,4.5]$. The y-axis displays values in logarithm scale. The minimum chance probability of $3\times10^{-7}$ is achieved with the set of parameters $L_{\rm min} = 4.0$ and $r_{\rm assoc}=0.55^{\circ}$. The estimated post-trial chance probability is $6\times10^{-7}$.}
\label{fig:significance}
\end{figure}

%
\begin{table*}[ht!] 
\vspace{2.5cm}
\begin{center}
\caption{Statistical significance of blazar/neutrino hotspots association. The chance probability of the correlation is estimated by performing $10^{8}$ MC simulations. The post-trial chance probability incorporates the effect of testing several datasets.}\label{tab:significance}
\begin{threeparttable}
\begin{tabular}{@{}lccccc@{}}
\toprule
Sky region & 5BZCat & Hotspots & Matches  & pre-trial p-value  & post-trial p-value\\
\hhline{======}
Southern sky ($L \geq 4$)     & 1177 & 19 & 10 &  $3 \times 10^{-7}$& $6\times 10^{-7}$ \\
\bottomrule
\end{tabular}
\end{threeparttable}
\end{center}
\end{table*}

At the end of this procedure, we obtain for each set of $(L^{i}_{\rm min},r^{j}_{\rm assoc})$ a corresponding pre-trial p-value, $p^{L_i,r_j}$ as shown in Figure \ref{fig:significance}. The minimum pre-trial p-value indicates the optimal set of $(L^{i}_{\rm min},r^{j}_{\rm assoc})$. In our analysis this corresponds to a  pre-trial p-value, $p^{\rm best}_{\rm pre}=3\times 10^{-7}$, achieved for $L_{\rm min}=4.0$ and $r_{\rm assoc}=0.55^{\circ}.$
This  minimum pre-trial p-value, obtained upon scanning the full space of parameter sets, provides us with the strongest potential correlation signal. As aforementioned, since these optimal values are derived from the data, a statistical penalty has to be evaluated and included in the calculation of the final chance probability, i.e. the post-trial p-value. The latter is evaluated using Eq.~1 following the same procedure that we have applied to the real data, as described above, with the difference that in place of TS$_{astro}^{L_i,r_j}$ we use each of the  TS$_{k}^{L_i,r_j}$ found from the MC samples. To this aim, we treat each MC sample as a ``real observation" and test it against the  $N=N_{\rm mock}-1$ random samples. This provides us with the total occurrences of p-values smaller or equal to $p^{\rm best}_{\rm pre}$ in the random MC population. This value is used as $M$ in Eq.~1, while $N=N_{\rm MC}-1$.
Following this approach we obtain a post-trial p-value $p^{\rm best}_{\rm post}=6\times 10^{-7}$, suggesting that the observed blazar/neutrino correlation is highly unlikely to arise by chance. Table \ref{tab:significance} summarises the outcomes of the statistical analysis.

%
%
\begin{figure*}[t!]
    \centering
    \includegraphics[scale=0.75]{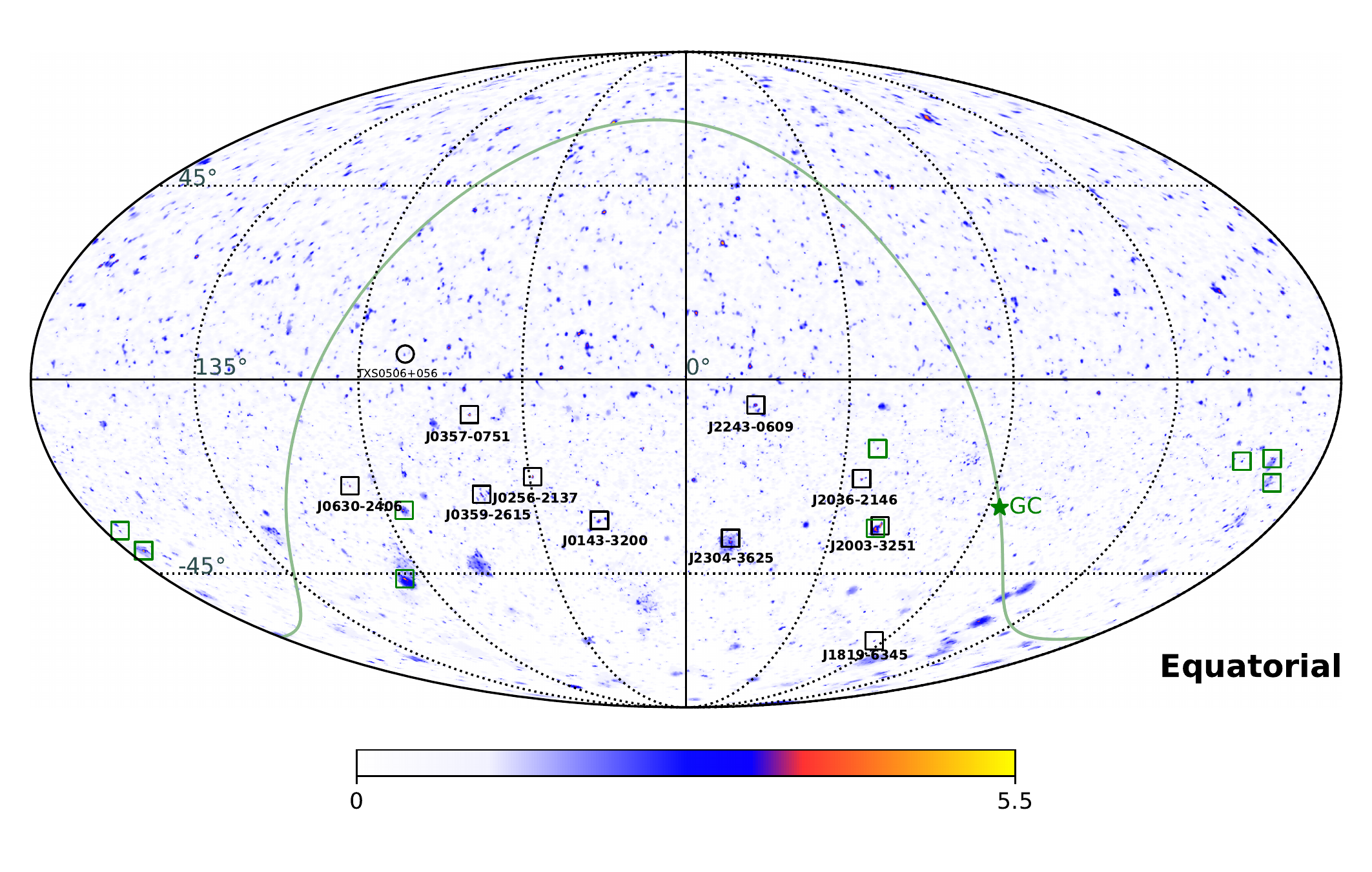}
    \caption{All-sky map in equatorial coordinates (J2000) of the IceCube neutrino local p-value logarithms denoted as $L$. Locations of PeVatron blazars associated with neutrino hotspots are pointed out by black squares. For visualization clarity, the label of 5BZCat objects is limited to report the unique numerical coordinate part. Unassociated hotspots are highlighted by green squares. The location of TXS~0506$+$056 is shown for reference (green circle). Squares are not to scale and serve the only purpose of highlighting  blazars' locations. %
    The Galactic plane and Galactic center are shown for reference as a green line and star, respectively.
    }
    \label{fig:allsky}
\end{figure*}
%

\section{PeVatron blazars hosted in the southern neutrino sky}
\label{sec:pevatrons}
Figure \ref{fig:allsky} displays the all-sky IceCube map in equatorial coordinates where we pinpoint the sky positions of the 5BZCat blazars associated with neutrino hotspots, i.e. the \emph{PeVatron blazars}. Among the southern neutrino sources associated with a BZCat object, with the exception of 5BZU~J1819$-$6345, all hotspot/blazar matches lie within declination $-40^{\circ} < \delta < -5^{\circ}$ i.e, close to the celestial horizon. This is the region with the best sensitivity of the IceCube Observatory, i.e. where a hard-spectrum astrophysical signal is more likely expected to emerge \citep{Coenders_PhDT:2016}. We note that despite being the blazar/neutrino association statistically robust, roughly half of the neutrino hotspots have no counterpart in 5BZCat. This may be a consequence of the incompleteness of the catalog \citep{massaro20155th}, as is generally the case for AGN catalogs. At the same time it opens to the possibility of other populations of neutrino emitters with e.g. different extragalactic and/or galactic counterparts at high latitudes \citep{IceCube10y:2020,Lhaaso:2021,LhaasoJ2108:2021}. Among the 9 unassociated hotspots, the majority have declination $\delta \gtrsim -40^{\circ}$ and thus are promising genuine astrophysical signals. Our result suggests that what we are observing may be the ``tip of the iceberg", i.e., the brightest most-efficient neutrino emitters. In fact, our statistical analysis focuses only on the handful of highest-significance hotspots. Due to the limited statistics, the optimal set of parameters identified in this work may be regarded as indicative values for the scales at which the correlation may be relevant. Following this pioneering discovery, more lower-confidence associations, hidden in the neutrino data and/or whose counterparts are not in 5BZCat, may be uncovered by more sophisticated techniques \citep{IceCube_AGNcores:2021} and more complete astrophysical population surveys.

\subsection{PeVatron Blazars are cosmic rays factories}
This work proves that at least part of the blazar population originates high-energy  neutrinos and, hence, is capable of accelerating cosmic rays.  A neutrino with observed energy $E$ must be produced at redshift $z$ with rest frame energy $E_{\nu}^{RF} = (1 + z)E$. If the neutrinos are produced by acceleration processes within the blazar jet, the relation between the rest-frame and observed energies is
$E_{\nu}^{RF} = (1 + z)E/D$, where $D=\frac{1}{\Gamma(1-\beta_{\Gamma} \cos(\theta))},$ is the beaming factor defined by the bulk Lorentz factor  $\Gamma$, and $\theta$ is the viewing angle of the jet. Typical beaming factors for blazars are of the order of $D\sim 10$ and  viewing angles are of the order of $\lesssim$ few degrees.
The production of neutrinos of energies E$_{\nu}$ unavoidably requires hadrons to be accelerated to energies $\sim 20 \times$~E$_{\nu}$~/~Z, Z being the atomic number \citep{Halzen:2013}. For the observed (minimum astrophysical) neutrino energies, i.e. between $\sim$100~TeV to $\sim$10~PeV energies, and assuming the acceleration of protons, the sample of PeVatron blazars diagnoses in situ acceleration of hadrons with energies above the PeV range. 

PeVatron blazars are listed in \autoref{tab:associations} along with the redshift information and neutrino hotspot association. The sample of PeVatron blazars presents a fairly large range of redshifts. The finding of a neutrino-emitter association within $\lesssim 280$~Mpc (\autoref{tab:associations}, z~=~0.063), i.e. not much beyond the GZK radius, tantalizes the connection of these newly discovered population of PeVatron blazars to the origin of UHECRs. As not many blazars may live within the GZK horizon, this implies that along with the blazar jet one may invoke as production site for the neutrinos additional scenarios such as neutrino emission from AGN-driven winds and/or disk-corona models. The excess of neutrinos in the direction of the misaligned jetted galaxy NGC~1068 reported by the IceCube collaboration \citep{IceCube10y:2020} and the detection of hadronic $\gamma$ rays from ultra fast outflows hosted in AGN \citep{UFO:2021} support this hypothesis.

On the opposite side, the objects found at high redshift open to the opportunity of studying the physics of powerful cosmic-ray accelerator sites at cosmological distances beyond the $\gamma$-ray cosmic horizon and the GZK horizon. The discovery of PeVatron blazars suggests that, in the cosmic-ray energy spectrum, the extragalactic component may contribute yet below the proton knee energies ($\sim$~PeV) putting into question the longstanding assumed postulate that the ankle ($\sim$3~EeV) marks the transition between Galactic and extragalactic cosmic rays. 

\section{Summary and Conclusions}
\label{sec:conclusions}
This analysis finds that 10 out of the 19 IceCube hotspots located in the southern sky are likely originated by blazars. We observe a roughly even distribution of neutrino hotspots across the southern sky. This corroborates the hypothesis that the dominant origin of these neutrino sources are blazars, which are isotropically distributed in the sky. The fact that half of the astrophysical-likely hotspots are associated with blazars fosters the idea that these newly discovered PeVatron blazars may be the dominant population of steady neutrino emitters resolved by IceCube at observed energies $E \gtrsim 100$~TeV.

It is important to put the discovery of PeVatron blazars in the context of recent works. Our findings are consistent with previous limits on the contribution by $\gamma$-ray blazars to the diffuse high-energy neutrino flux observed by IceCube \citep{IceCube2017_2LAC,IceCube-Gen2_WP:2021,Yuan:2020}, being only a small fraction ($\lesssim 30\%$) of the neutrino-emitter blazars detected also at GeV $\gamma$ rays (see also Appendix \ref{appendix:previous_gamma}).
This suggests that in the blazars' engine the neutrino emission is weakly related to the observed $\gamma$-ray emission. This implies different production sites for the bulk of the observed neutrinos and GeV $\gamma$ rays in  blazars \citep{Murase2016_gamma_dark,Reimer2019}. The outcomes presented in this paper are in agreement with the conclusions presented by \citet{Murase2016_gamma_dark}, that $\gamma$-ray-weak blazars may harbor efficient cosmic-ray accelerators able to produce $\sim$PeV neutrinos, motivating to explore physical models with predictions in the X-ray and MeV spectral range. 

Our finding indicates a firm indirect detection of extragalactic cosmic-ray factories with \emph{in situ} acceleration of cosmic rays to PeV energies and, possibly, up to the EeV regime (assuming the acceleration of protons). 
PeVatron blazars shed a new perspective in the properties of the cosmic-ray spectrum, as well as offer a promising probe to test fundamental particle-physics properties beyond the energy region accessible by LHC. The non-detection of individual $\gtrsim 10$~PeV likely-astrophysical neutrinos over a decade of IceCube observations \citep{IceCube_Glashow:2021} may imply a physical intrinsic limit for PeVatron blazars, i.e. related to the maximum energy of the parent cosmic rays. Nonetheless, the lack of statistics above tens of PeV could be simply due to the sensitivity of IceCube that at those energies degrades rapidly. In the latter case, PeVatron blazars may accelerate hadrons to much higher energies, fostering the tantalizing prospect that the observed high-energy astrophysical neutrinos and UHECRs could be produced by the same population of cosmologically distributed sources \citep{Waxmal_Cosmological:2014,Murase_blazarUHECRs:2012}. The forthcoming generation of new neutrino detectors such as IceCube-Gen2 \citep{IceCube-Gen2_WP:2021}, the Cubic Kilometre Neutrino Telescope \citep[KM3NeT,][]{KM3NeT:2016}, the The Pacific Ocean Neutrino Experiment \citep[P-ONE,][]{P-ONE:2020}, the Radio Neutrino Observatory in Greenland \cite[RNO-G,][]{RNO-G:2021} and the Giant Radio Array for Neutrino Detection \citep[GRAND,][]{GRAND:2020} project has the potential of shedding light into this.

%
%
\begin{table*}
\begin{center}
\caption{List of PeVatron blazars / neutrino hotspot associations found by this work. The columns report the neutrino hotspots equatorial coordinates (J2000) and the $L$ value of the hotspot. The candidate 5BZCat blazar counterpart along with the redshift and the distance between the blazar and hotspot. In the lower part of the table we report for reference the hotspots without a 5BZCat association.}\label{tab:associations}
\begin{threeparttable}
\resizebox{0.9\textwidth}{!}{ 
\begin{tabular}{@{}lrcclcc@{}}
\toprule
IceCube hotspots &&&& Blazar associations\\
\hhline{=======}
 & $\alpha_{hs}$[$^{\circ}$] & $\delta_{hs}$[$^{\circ}$] & $L$  & 5BZCat & z &  Separation[$^{\circ}$]\\ 
\cline{1-7} 
\object[IC J2243$-$0540]{IC J2243$-$0540}  &  340.75  &  $-$5.68   &  4.012 & \object[5BZB J2243$-$0609]{5BZB J2243$-$0609}     &  0.30$^c$         & 0.47 \\ 
 \object[IC J0359$-$0746]{IC J0359$-$0746} &  59.85   &  $-$7.78   &  5.565 & \object[5BZQ J0357$-$0751]{5BZQ J0357$-$0751}     &  1.05             & 0.42 \\ 
 \object[IC J0256$-$2146]{IC J0256$-$2146} &  44.12   &  $-$21.78  &  4.873 & \object[5BZQ J0256$-$2137]{5BZQ J0256$-$2137}     &  1.47             & 0.17 \\ 
 \object[IC J2037$-$2216]{IC J2037$-$2216} &  309.38  &  $-$22.27  &  4.664 & \object[5BZQ J2036$-$2146]{5BZQ J2036$-$2146}     &  2.299            & 0.51 \\ 
 \object[IC J0630$-$2353]{IC J0630$-$2353} &  97.56   &  $-$23.89  &  4.420 & \object[5BZB J0630$-$2406]{5BZB J0630$-$2406$^{a,b}$}& $>$1.238$^{d}$ & 0.28 \\ 
 \object[IC J0359$-$2551]{IC J0359$-$2551} &  59.94   &  $-$25.86  &  4.356 & \object[5BZB J0359$-$2615]{5BZB J0359$-$2615$^a$} & 1.47$^{e}$        & 0.40 \\ 
 \object[IC J0145$-$3154]{IC J0145$-$3154} &  26.28   &  $-$31.91  &  4.937 & \object[5BZU J0143$-$3200]{5BZU J0143$-$3200$^a$} &  0.375            & 0.42 \\ 
 \object[IC J2001$-$3314]{IC J2001$-$3314} &  300.41  &  $-$33.24  &  4.905 & \object[5BZQ J2003$-$3251]{5BZQ J2003$-$3251}     &  3.773            & 0.53 \\ 
 \object[IC J2304$-$3614]{IC J2304$-$3614} &  346.03  &  $-$36.24  &  4.025 & \object[5BZQ J2304$-$3625]{5BZQ J2304$-$3625}     &  0.962            & 0.24 \\ 
 \object[IC J1818$-$6315]{IC J1818$-$6315} &  274.50  &  $-$63.26  &  4.030 & \object[5BZU J1819$-$6345]{5BZU J1819$-$6345}     &  0.063            & 0.53 \\ 
\midrule
\object[IC J2024$-$1524]{IC J2024$-$1524}  &  306.12 &  $-$15.40   & 4.454 & --  &  --  & --\\ 
\object[IC J1256$-$1739]{IC J1256$-$1739}  &  194.06 &  $-$17.66   & 4.407 & --  &  --  & --\\ 
\object[IC J1329$-$1817]{IC J1329$-$1817}  &  202.32 &  $-$18.29   & 4.040 & --  &  --  & --\\ 
\object[IC J1241$-$2314]{IC J1241$-$2314}  &  190.37 &  $-$23.24   & 4.288 & --  &  --  & --\\ 
\object[IC J0538$-$2934]{IC J0538$-$2934}  &   84.73 &  $-$29.57   & 4.994 & --  &  --  & --\\ 
\object[IC J2006$-$3652]{IC J2006$-$3352  &  301.55 &  $-$33.87}  & 4.698 & --  &  --  & --\\ 
\object[IC J1140$-$3424]{IC J1140$-$3424}  &  175.17 &  $-$34.41   & 4.082 & --  &  --  & --\\ 
\object[IC J1138$-$3915]{IC J1138$-$3915}$^{f}$  &  174.64 &  $-$39.26   & 5.885 & --  &  --  & --\\ 
\object[IC J0628$-$4616]{IC J0628$-$4616}  &  97.23  &  $-$46.28   & 4.987 & --  &  --  & --\\ 
\botrule
\end{tabular}}
\begin{tablenotes}
    \begin{minipage}{0.9\textwidth}
    \vspace{0.1cm}
	\small 
	\item
$^{a}$ Blazars listed as $\gamma$-ray emitters in 4FGL-DR2 \citep{ballet2020_4FGL-DR2}.
$^{b}$ Blazar listed in 2LAC \citep{2LAC}.
$^{c}$ Redshift from \citet{Crates:2007}. 
$^{d}$ Redshift from \citet{Shaw:2013}.
$^{e}$ Redshift from \citet{Drinkwater:1997}.
$^{f}$ Hotspot positionally consistent with a neutrino clustering excess previously identified in the 7-year dataset \citep[][]{IceCube7y:2017} and 3-year dataset \citep{Smith:2021}.
    \end{minipage}
\end{tablenotes}
\end{threeparttable}
\end{center}
\end{table*}

\begin{acknowledgments}
This work was supported by the European Research Council, ERC Starting grant \emph{MessMapp}, S.B. Principal Investigator, under contract no. 949555. 
S.B. and A.T. are grateful for valuable conversation to M. Santander, K. Murase, M. Petropoulou, J. DeLaunay, G. Illuminati, D. Caprioli, D. Bastieri, R. D'Abrusco, M. Giroletti, A. Maselli, F. Massaro, S. I. Stathopoulos, A. Kouchner. 
This work has made use of data from the Space Science Data Center (SSDC), a facility of the Italian Space Agency (ASI), and data provided by the IceCube Observatory. 
\end{acknowledgments}

\facilities{The IceCube Observatory}
\software{astropy \citep{2013A&A...558A..33A,2018AJ....156..123A},  
          healpy, HEALPix, topcat
          }

\appendix

\section{Statistical analysis robustness}\label{sec:appendix_statistics}
\subsection{Accessing potential bias}\label{appendix:scan}
Given the IceCube location at the South Pole, the effective area changes with declination and energy, therefore the expected background in the IceCube data depends strongly on the declination. For the 7-years skymap provided by \citet{IceCube7y:2017} the final p-values reported in each sky-pixel have been trial-corrected accounting for the chance of background fluctuations occurring at any position in the sky. The skymap p-values, that represent the neutrino clustering probabilities (i.e., $L$-values), obtained in such manner are thus declination-independent \citep{IceCube7y:2017}.

\begin{figure}
  \centering
    \includegraphics[scale=0.45]{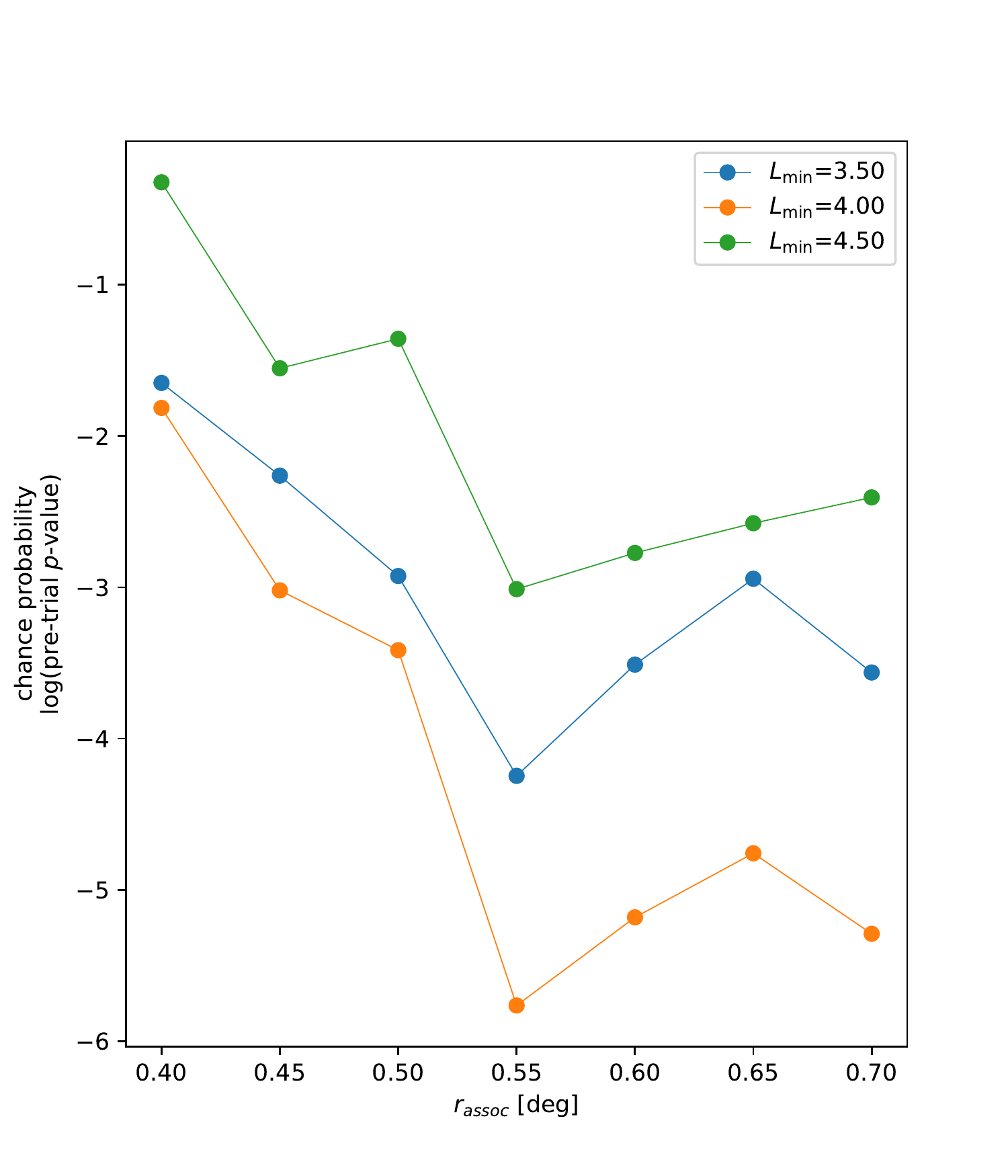} 
    \caption{Same as Fig. \ref{fig:significance} but obtained by randomising the neutrino spot positions in right ascension (see Appendix \ref{appendix:scan}). Pre-trial p-value for the blazar/neutrino correlation as a function of the association radii $r_{\rm assoc}$, for the neutrino dataset with $L_{\rm min}=[3.5,4.0,4.5]$. The y-axis displays values in logarithm scale. The minimum chance probability of $1.7\times10^{-6}$ is achieved with the set of parameters $L_{\rm min} = 4.0$ and $r_{\rm assoc}=0.55^{\circ}$. The estimated post-trial chance probability is $1.8\times10^{-6}$.}
    \label{fig:HS_ra}
  \end{figure}

As a-posteriori, further robustness test we repeated the cross-correlation analysis by keeping the blazar sky distribution fixed and randomizing the IceCube neutrino spots. This was done by randomizing the Right Ascension of the neutrino spots while keeping their declination fixed, thus accounting for the potential, unaccounted declination dependence in the IceCube acceptance, and repeating the steps of the statistical analysis described in the main text. This procedure yields a minimum pre-trial p-value of $1.7 \times 10^{-6}$ which is achieved for $L_{\rm min}=4.0$ and $r_{\rm assoc}=0.55^{\circ}$, and a post-trial p-value of $1.8 \times 10^{-6}$ (see Figure \ref{fig:HS_ra}). A slighter higher post-trial p-value with respect to the one in Table \ref{tab:significance} is expected since in these experiments the randomization is applied only to the Right Ascension.

\subsection{Null hypothesis statistical properties}\label{appendix:ks}
The statistical positional correlation analysis presented in Section \ref{sec:cross-corr} performs $10^8$ simulations, implemented as $10^2 \times 10^6$ mocks for computational reasons, where we randomly shift the sky position of the original 5BZCat sources by a random value between 0 and 10 degrees. In the randomization we ensure that the angular distribution of the catalog is preserved performing Kolmogorov-Smirnov (KS) tests on the distributions of the Right Ascension  and declinations between the real (5BZCat) and mock catalog. For each simulation of the $10^6$ run we record the corresponding KS p-values. Once the full run is completed we record the minimum KS p-value among them, for both the RA and Dec distributions. This process is repeated $10^2$ times, in order to obtain $10^8$ simulations. The total number of $10^8$ simulations to be performed is imposed by the requirement of observing an equal number or more of matches as observed in the real data ($\geq 10$ matches) in at least one mock simulation (frequentist approach).

In the 5BZCat randomization procedure, the majority (95\% quantile) of the minimum KS p-values obtained for the mocks versus real data distributions have minimum KS p-values larger than $\sim0.1$ and $\sim0.07$ for RA and Dec, respectively. The lowest minimum p-value recorded is 0.02. Such KS p-values ensure that the mock sky distributions are compatible with the original distribution, i.e. the mocks distributions obtained in this manner preserve the statistical properties of the null hypothesis.
The same validation is performed on the  a-posteriori neutrino hotspot randomisation (see Appendix \ref{appendix:scan}) yielding consisting results.

%
%

\section{Comparison to all-sky neutrino hotspot population searches}\label{appendix:allsky}
The non-observation of individual neutrino sources, i.e. hotspots with $L>>6$, in the previous searches is not in conflict with our results. Neutrino hotspots represent the level of spatial clustering in the neutrino data at a given sky location. Therefore, none of the neutrino hotspots, and consequently blazars associated, may necessarily be detected at high confidence by IceCube.
Besides, the all-sky scan performed by e.g.~\citet{IceCube7y:2017} and similar searches \citep[][]{Hooper:2019,Smith:2021} requires the astrophysical signal hypothesis to exceed the highest background fluctuations from many trials in order to be significant. Implying that, although the IceCube all-sky analysis is sensitive to an astrophysical signal from any direction, the source requires a strong flux to be significantly detected individually in the all-sky scan \citep[][]{Coenders_PhDT:2016}. Even if each neutrino source is not individually detected, the statistical cross-correlation technique presented here has the capability of pinpointing a correlation between sub-threshold genuine-astrophysical neutrino emitters and astrophysical sources. In principle more than one astrophysical object may contribute to enhancing the neutrino clustering local p-value for a spot.
Accordingly, our finding claims neither that all 5BZCat sources are neutrino emitters nor that the PeVatron blazars emit neutrinos at the same level. It rather unravels a sub-population of blazars that may host more efficient hadronic accelerators at $\sim$PeV energies.

\section{Comparison to previous works on gamma-ray-selected blazars}\label{appendix:previous_gamma}
Previous works applied a neutrino stacking analysis to infer limits on the cumulative neutrino emission from $\gamma$-ray blazars listed in the 2nd $Fermi$-LAT AGN Catalog \citep[2LAC,][]{2LAC}. Results from \citet{IceCube2017_2LAC} showed that 2LAC blazars contribute for $\lesssim 27\%$ to the diffuse neutrino flux between 10~TeV and 2~PeV, when assuming a power-law spectrum with spectral index of $-2.5$. The constraint weakens \citep{aartsen2018_neutrino_prior_to_TXS} to about $40\% - 80\%$ of the total observed neutrino flux assuming a spectral index of $-2$, as it is observed for the astrophysical diffuse flux at the higher energies \citep[$\gtrsim 200$~TeV,][]{IC_north_hard_spectrum:2016}.
Among the 10 neutrino-blazar associations found by our work and listed in Table \ref{tab:associations}, only one was reported as $\gamma$-ray emitter in the 2LAC and hence included in the stacking analysis. Thus, our finding is consistent with previous neutrino stacking limits reported by the IceCube collaboration.

We note that three of the associated blazars have a counterpart at $\gamma$ rays in the 4th $Fermi$-LAT Catalog \citep[4FGL-DR2,][]{ballet2020_4FGL-DR2}, suggesting that in future more neutrino-emitter blazars may be detectable also at $\gamma$ rays. The newly discovered PeVatron blazars may represent an emerging population of weak $\gamma$-ray emitters, potentially detectable with  larger accumulation of statistics with the \textit{Fermi}-LAT  and/or the enhanced sensitivity of Cherenkov telescopes.
However, the lack of  $\gamma$-ray counterparts for most of the associations in Table \ref{tab:associations} supports the idea that the $\gamma$-ray brightness of a blazar in the \textit{Fermi}-LAT energy band (GeV energies) may be not necessarily correlated with the neutrino brightness, in agreement with \citet{Murase2016_gamma_dark}.

\section{Comparison to previous neutrino/blazars correlation studies}
Conflicting results exist regarding a possible link between the radio brightness of blazars and putative neutrino emission \citep{plavin2021directional,Hovatta:2021,Bei_noradio_corr:2021}. The 5BZCat catalog includes archival information about the integrated-radio flux in the bands 1.4~GHz and 5~GHz for each object. We do not observe any noteworthy difference in the  integrated-radio flux in the bands 1.4~GHz and 5~GHz for the neutrino-emitter blazars found by our work and the other 5BZCat sources. This is consistent with findings of \citet{Hovatta:2021,Bei_noradio_corr:2021}, which report a low level of correlation (chance probability $\gtrsim 0.6\%$) between AGN selected at these frequencies and IceCube neutrinos.
For reference, the radio-selected sample from \citet{plavin2021directional} has 715 objects included in 5BZCat. Since that study \citep{plavin2021directional} focuses on the northern hemisphere only, no direct comparison with our work can be drawn.

Another study by \citet{luo2020} reports no evidence for a correlation between the 5BZCat sources and a sample of 45 IceCube high-energy neutrino events. The main difference with the work presented here is the neutrino dataset. The former study uses a small number (45) of individual neutrino high-energy events. We make use of an all-sky neutrino map that exploits a large statistics of neutrino events accumulated over 7 years. The all-sky map is built with a refined likelihood analysis where the neutrino data are optimized for searches of point-like neutrino sources with hard powerlaw spectrum. 

A search for a correlation between 5BZCat objects and IceCube neutrinos by \citet{Hooper:2019} led to inconclusive results. The main difference is that \citet{Hooper:2019} utilises only one year of neutrino data (the 86-string data), while in our work we employ a larger dataset collected over a longer period of time. 

\section{TXS 0506+056: a promising PeVatron blazar}
Based on our work one may predict that the IceCube observatory will reach the sensitivity to detect individual astrophysical point-sources at high confidence in the near future. This behavior is yet observed at the location of \TXS, associated with the 5BZCat object 5BZB~J0509+0541, and has been claimed to be a neutrino-emitter blazar \citep{icecube2018_TXS_flaring}. In the 7-year IceCube data utilised by this work, it appears in spatial agreement with a neutrino spot of $L = 2.2$. Since it is located in the northern hemisphere, this blazar is not included in our statistical analysis. However, we note that in the analysis of 10 (8) years of IceCube observations \citep{IceCube10y:2020,IceCube8yNorth:2019}, i.e. 3 (2) additional years compared to the all-sky map used by us, the value of $L$ in coincidence with \TXS\ progressively increases to 3.72 (2.65), as expected for a truly astrophysical signal that keeps steadily increasing when deepening the observational sensitivity and acquiring more exposure. Besides, \citet{IceCube10y:2020} reports that the cumulative 10-year signal at the location of \TXS\ is best-fitted by a hard powerlaw ($\propto E^{-2.1}$) neutrino spectrum, that is consistent with predictions of blazar hadronic models. This corroborates the hypothesis that this blazar may be a genuine astrophysical neutrino source. It would be interesting applying our analysis to the IceCube 10-year all-sky likelihood map, which has not been released publicly at the time of the writing.

\bibliography{main}{}
\bibliographystyle{aasjournal}
\end{document}